\documentclass[11pt]{article}
\usepackage{amsmath}

\newcommand{\bra}[1]{\mbox{$\langle\ {#1}\ |$}}
\newcommand{\ket}[1]{\mbox{$|\ {#1}\ \rangle$}}
\newcommand{\braket}[2]{\mbox{$\langle\ {#1}\ |\ {#2}\ \rangle$}}

\newtheorem{theorem}{Theorem}

\begin{document}
\title{\bf SPECTRAL DECOMPOSITION OF AN ELEMENTARY 3-FERMION 2-BODY
OPERATOR}
\author{Hubert Grudzi\'nski\\
Department of Physics, Academy of Bydgoszcz\\
85--072 Bydgoszcz, pl. Weyssenhoffa 11, Poland\\
(e-mail: hubertg@ab-byd.edu.pl)\\
Jacek Hirsch\\
Institute of Physics, Nicholas Copernicus University\\
87--100 Toru\'n, Poland\\
(e-mail: jacekh@phys.uni.torun.pl)}
\date{}
\maketitle
\begin{abstract}
The eigenvalues and eigenfunctions of an elementary 3-fermion
2-body operator $3P^2_g\wedge I^1\equiv A^3 \sum\limits_{1\leq i <
j \leq 3 } P^2_g(i,j)A^3$ acting on a 3-particle antisymmetric
finite dimensional Hilbert space have been found. Here $P^2_g$
denotes the projection operator onto a 2-particle antisymmetric
function $g^2$, while $A^3$ denotes the 3-particle
antisymmetrizing operator.
\end{abstract}
{\bf keywords: spectral decomposition of operators, fermion 2--body operators}
%
%
%
\section{Introduction}

The spectral decomposition of operators is an interesting subject in its own right.
This paper arose while we were searching for new conditions for
fermion $N$-representability
\cite{Coleman63}--\cite{GarrodPercus64}, \cite{Kummer67,
Kummer77}. The new condition for fermion $N$-representability, the
"dual $P$-condition" \cite{GrudzinskiHirsch}, requires a knowledge
of the maximal eigenvalue of the positive semidefinite operator
${N \choose 2} P^2_g\wedge I^{\wedge (N-2)}$ acting on an
$N$-particle antisymmetric Hilbert space ${\cal H}^{\wedge N}$
(the $N$-fold Grassmann product of ${\cal H}^1$), where $P^2_g$ is
the projection operator onto a 2-particle antisymmetric function
$g^2\in {\cal H}^{\wedge 2}$, and $I^{\wedge (N-2)}$ denotes the
identity operator on ${\cal H}^{\wedge (N-2)}$. We call this
operator an elementary $N$-fermion 2-body operator ("with 2-body
interactions"). So far, we were able to find the spectral
decomposition of such an operator for arbitrary $g^2$ only for
$N=3$, and this paper contains the results. It is realistic to
solve the problem for arbitrary $N$ if $g^2$ is of a special type,
e.g. "extreme geminal" \cite{Coleman65}, and we will publish those
results later. Having the spectral decomposition of an elementary
$N$-fermion 2-body operator it is possible to find the reduced
2-particle density operators corresponding to its eigenfunctions
and thus obtain some detailed information about the structure of
the convex set ${\cal P}^2_N$ consisting of all 2-particle fermion
$N$-representable density operators. Especially, the reduced
2-density operators corresponding to the kernel (null-space) of
the operator ${N \choose 2} P^2_g\wedge I^{\wedge (N-2)}$ are
interesting because they form a face in the set ${\cal P}^2_N$
exposed by the operator $P^2_g$ \cite{Kummer67}. We have these
results for $N=3$ (they will be published in a separate paper),
and for this purpose we have introduced in this paper in the
null-space of the operator $3P^2_g\wedge I^1$ an orthonormal
basis, and give explicitly the projection operator onto the
kernel. Theorem 1 contains the spectral decomposition of the
operator $3P^2_g\wedge I^1$, while Theorem 2 gives the projection
operator ${\rm Ker}\,(3P^2_g\wedge I^1)$ (upper-case K) onto the
null-space $\ker(3P^2_g\wedge I^1)$ (lower-case k) of this
operator.
%
%
%
%
\section{Spectral decomposition}

\begin{theorem}
Let ${\cal H}^1$ be a finite dimensional Hilbert space $(\dim
{\cal H}^1=n)$, and ${\cal H}^{\wedge 2}\equiv {\cal H}^1\wedge
{\cal H}^1$ denotes the 2-particle antisymmetric space generated
by ${\cal H}^1$ (the 2-fold Grassmann product of ${\cal H}^1$).
Let $P^2_g$ denote the 1-dimensional projection operator onto a
2-particle antisymmetric function $g^2\in {\cal H}^{\wedge 2}$ of
1-rank $r=2s$ possessing the canonical decomposition
$g^2=\sum^s_{i=1}\xi_i \ket{2i-1,2i}$ with
$\sum^s_{i=1}|\xi_i|^2=1$, where $\ket{2i-1,2i}\equiv
\sqrt{2}\,\varphi^1_{2i-1}\wedge \varphi^1_{2i}\equiv
\frac{1}{\sqrt{2}} \det(\varphi^1_{2i-1},\varphi^1_{2i})$,
$\varphi^1_i\in {\cal H}^1$, is the 2-particle normalized
determinant. Let the identity operator $I^1$ on ${\cal H}^1$
possess the decomposition
$I^1=\sum^{r=2s}_{i=1}P^1_i+\sum^n_{i=r+1}P^1_i$, where
$P^1_i\equiv \varphi^1_i\otimes\overline{\varphi}^1_i\equiv
\ket{i}\bra{i}$ $(i=1,\dots,n)$ are 1-dim mutually orthogonal
projection operators onto the 1-particle functions
$\ket{i}\equiv\varphi^1_i \in{\cal H}^1$ $(i=1,\dots,n)$ forming
an orthonormal basis in ${\cal H}^1$. Then, the 3-particle
operator $3P^2_g\wedge I^1:{\cal H}^{\wedge 3}\longrightarrow
{\cal H}^{\wedge 3}$ possesses the following spectral
decomposition
\begin{eqnarray}
3P^2_g\wedge I^1=\sum^{s=r/2}_{k=1} \big(1-|\xi_k|^2\big)
\Big(P^3_{g_{2k-1}} + P^3_{g_{2k}}\Big) + \sum^n_{l=r+1}
P^3_{g_l}+0\cdot{\rm Ker\,}(3P^2_g\wedge I^1). \label{sd}
\end{eqnarray}
Here, $P^3_{g_{2k-1}}$, $P^3_{g_{2k}}$ $(k=1,\dots,s=r/2)$,
$P^3_{g_l}$ ($l=r+1,\dots,n$) are 1-dim projectors onto the
following eigenfunctions:

\begin{eqnarray}
\label{ef1}
g^3_{2k-1}&=&\frac{1}{\sqrt{1-|\xi_k|^2}}\sum^s_{i=1\atop (i\ne
k)} \xi_i\ket{2i-1,2i,2k-1}=\sqrt{\frac{3}{1-|\xi_k|^2}}\
g^2\wedge \varphi^1_{2k-1},\\
&& \hspace*{9cm}\quad (k=1,\dots,s=\frac{r}{2})\nonumber
\end{eqnarray}
\begin{eqnarray}
g^3_{2k}&=&\frac{1}{\sqrt{1-|\xi_k|^2}}\sum^s_{i=1\atop (i\ne k)}
\xi_i\ket{2i-1,2i,2k}=\sqrt{\frac{3}{1-|\xi_k|^2}}\ g^2\wedge
\varphi^1_{2k},\label{ef2}
\end{eqnarray}
\begin{eqnarray}
g^3_l=\sum^s_{i=1} \xi_i \ket{2i-1,2i,l}=\sqrt{3}\ g^2\wedge
\varphi^1_l, \quad (l=r+1,\dots,n), \label{ef3}
\end{eqnarray}
while ${\rm Ker\,}(3P^2_g\wedge I^1)$ denotes the projection
operator onto the null-space $\ker(3P^2_g\wedge I^1)$ of the
operator $3P^2_g\wedge I^1$, which is of dimension ${n \choose
3}-n$. The symbols of the type $\ket{2i-1,2i,2k-1}\equiv
\sqrt{3!}\,\varphi^1_{2i-1}\wedge \varphi^1_{2i} \wedge
\varphi^1_{2k-1}$ denote the appropriate 3-particle normalized
determinants.
\end{theorem}
%
%
%
%
{\bf Proof.} First, we observe that the functions defined by
(\ref{ef1}--\ref{ef3}) are normalized and all are mutually
orthogonal, because the determinants differ from each other in
at least one 1-particle function. To prove that they are the
eigenfunctions belonging to non-zero eigenvalues of the operator
$3P^2_g\wedge I^1$, we express $P^2_g$ and $I^1$ in the following
way:
\begin{eqnarray*}
&&P^2_g=\sum^{s=r/2}_{i,j=1} \xi_i\overline\xi_j
\ket{2i-1,2i}\bra{2j-1,2j}\\
&&I^1=\sum^{s=r/2}_{k=1}\Big(P^1_{2k-1}+P^1_{2k}\Big)+\sum^n_{l=r+1}P^1_l=
\sum^s_{k=1}\big(\ket{2k-1}\bra{2k-1}+\ket{2k}\bra{2k}\big)+\sum^n_{l=r+1}\ket{l}\bra{l}.
\end{eqnarray*}
Then,
\begin{eqnarray*}
3P^2_g\wedge I^1 &=& 3\sum^s_{i,j=1}\sum^s_{k=1}
\xi_i\overline\xi_j \ket{2i-1,2i}\bra{2j-1,2j}
\wedge\big( \ket{2k-1}\bra{2k-1}+\ket{2k}\bra{2k}\big) +\\
&+&3\sum^s_{i,j=1}\sum^n_{l=r+1}\xi_i\overline\xi_j
\ket{2i-1,2i}\bra{2j-1,2j}\wedge\ket{l}\bra{l}=\\
&=&\sum^s_{k=1}\big(1-|\xi_k|^2\big) \left(
\frac{1}{1-|\xi_k|^2} \sum^s_{i,j=1\atop (i,j\ne k)}
\xi_i\overline\xi_j \ket{2i-1,2i,2k-1}\bra{2j-1,2j,2k-1}+\right. \\
&+&\left.\frac{1}{1-|\xi_k|^2} \sum^s_{i,j=1\atop (i,j\ne k)}
\xi_i\overline\xi_j \ket{2i-1,2i,2k}\bra{2j-1,2j,2k}\right)+\\
&+&\sum^n_{l=r+1}\left(\sum^s_{i,j=1}\xi_i\overline\xi_j
\ket{2i-1,2i,l}\bra{2j-1,2j,l}\right)=\\
&=&\sum^s_{k=1}\big(1-|\xi_k|^2\big)\Big(\ket{g^3_{2k-1}}\bra{g^3_{2k-1}}+
\ket{g^3_{2k}}\bra{g^3_{2k}}\Big)+\sum^n_{l=r+1}
\ket{g^3_l}\bra{g^3_l}=\\
&=&\sum^s_{k=1}\big(1-|\xi_k|^2\big)
\Big(P^3_{g_{2k-1}}+P^3_{g_{2k}} \Big)+\sum^n_{l=r+1}P^3_{g_l}.
\end{eqnarray*}
In the proof we have used the fact that
$\sqrt{3}\,\ket{2i-1,2i}\wedge\ket{k}=\ket{2i-1,2i,k}\equiv\sqrt{3!}\,
\varphi^1_{2i-1}\wedge\varphi^1_{2i}\wedge\varphi^1_k\equiv\det(
\varphi^1_{2i-1},\varphi^1_{2i},\varphi^1_k)$. The symbol of the
type $\ket{g^3_{2k-1}}\bra{g^3_{2k-1}}\equiv g^3_{2k-1}\otimes
\overline g^3_{2k-1} \equiv P^3_{g_{2k-1}}$ denotes the projection
operator onto the function $g^3_{2k-1}\in {\cal H}^{\wedge 3}$,
while $\ket{2i-1,2i,k}\bra{2i-1,2i,k}$ is the projection operator
onto the determinant function $\ket{2i-1,2i,k} \in {\cal
H}^{\wedge 3}$.

Since the 1-dimensional projectors $P^3_{g_{2k-1}}$,
$P^3_{g_{2k}}$, $P^3_{g_l}$ are mutually orthogonal, the obtained above 
expression
\begin{eqnarray*}
3P^2_g\wedge I^1=\sum^s_{k=1}\big(1-|\xi_k|^2\big)
\Big(P^3_{g_{2k-1}}+P^3_{g_{2k}} \Big)+\sum^n_{l=r+1}P^3_{g_l}
\end{eqnarray*}
is the spectral resolution of the operator $3P^2_g\wedge I^1$
corresponding to the $n$ non-zero eigenvalues. The orthogonal
complement in ${\cal H}^{\wedge 3}$ of the subspace
%
%
%
%
spanned by the eigenfunctions $g^3_{2k-1}$, $g^3_{2k}$
($k=1,\dots,s=r/2$), $g^3_l$ ($l=r+1,\dots,n$) is the null-space
(kernel) of the operator $3P^2_g\wedge I^1$ corresponding to the
eigenvalue zero. The projection operator onto this ${n \choose
3}-n$ dimensional null-space we denote by ${\rm Ker}\,( 3P^2_g
\wedge I^1)$. Thus, we have obtained resolution (\ref{sd}), and
this completes the proof.

Since we are interested in the reduced density operators
corresponding to the eigenfunctions belonging to the eigenvalue
zero of the operator  $3P^2_g\wedge I^1$, we have introduced an
orthonormal basis in the null-space $\ker( 3P^2_g\wedge I^1)$ and
have found the projection operator ${\rm Ker}\,( 3P^2_g \wedge
I^1)$ onto this null-space explicitly.

\begin{theorem}
The projection operator ${\rm Ker}\,( 3P^2_g \wedge I^1)$ onto the
null-space of the operator $3P^2_g \wedge I^1$ $(g^2 \in{\cal
H}^{\wedge 2})$ is a sum of mutually orthogonal projectors
\begin{eqnarray}
{\rm Ker}\,( 3P^2_g \wedge
I^1)=K^3_{0,3}+K^3_{1,2}+K^3_{2,1}+K^3_{3,0}, \label{kd0}
\end{eqnarray}
corresponding to the orthogonal decomposition of the 3-particle
antisymmetric Hilbert space
\begin{eqnarray}
{\cal H}^{\wedge 3}=\widetilde {\cal R}^{\wedge 3}\oplus {\cal
R}^1\wedge \widetilde{\cal R}^{\wedge 2} \oplus {\cal R}^{\wedge
2}\wedge \widetilde{\cal R}^1\oplus{\cal R}^{\wedge 3},
\end{eqnarray}
with ${\cal H}^1={\cal R}^1\oplus \widetilde{\cal R}^1$, where
${\cal R}^1$ denotes the subspace spanned by the orthonormal basis
$\{\varphi^1_i \}^r_{i=1}$, and $\widetilde {\cal R}^1$ by
$\{\varphi^1_i \}^n_{i=r+1}$. The projection operators
$K^3_{0,3}$, $K^3_{1,2}$, $K^3_{2,1}$, $K^3_{3,0}$ can be
expressed in the following form:
\begin{eqnarray}
K^3_{0,3}&=&\sum_{r+1 \leq j_1<j_2<j_3\leq n} P^3_{j_1,j_2,j_3},\\
K^3_{1,2}&=&\sum^s_{i=1}\sum_{r+1\leq j_1<j_2\leq n} \Big(
P^3_{2i-1,j_1,j_2}+P^3_{2i,j_1,j_2} \Big),\\
K^3_{2,1}&=&\sum_{1\leq i_1<i_2\leq s}\ \sum_{j=r+1}^n \Big(
P^3_{2i_1-1,2i_2-1,j} + P^3_{2i_1,2i_2,j} + P^3_{2i_1-1,2i_2,j}+
P^3_{2i_1,2i_2-1,j}
 \Big)+\nonumber\\
&+&\sum^n_{l=r+1}\sum^s_{m=2}P^3_{f_{l,m}},\\
K^3_{3,0}&=&\sum_{1\leq i_1 < i_2 < i_3 \leq s}
\Big(P^3_{2i_1-1,2i_2-1,2i_3-1}+ P^3_{2i_1-1,2i_2-1,2i_3}+
P^3_{2i_1-1,2i_2,2i_3-1}+\nonumber\\
&+&P^3_{2i_1,2i_2-1,2i_3-1}+P^3_{2i_1-1,2i_2,2i_3}+
P^3_{2i_1,2i_2-1,2i_3}+ P^3_{2i_1,2i_2,2i_3-1}
+P^3_{2i_1,2i_2,2i_3}\Big)+ \nonumber\\
&+&\sum^s_{k=1}\sum_{m\in J}\Big(P^3_{f_{2k-1,m}}+
P^3_{f_{2k,m}}\Big), \label{kd1-4}
\end{eqnarray}
%
%
%
%
\[
J=\left\{
\begin{array}{l@{\quad \mbox{for} \quad}l}
\{3,\dots,s\},& k=1\\
\{2,\dots,k-1,k+1,\dots,s\},& k=2,\dots,s.
\end{array}\right.
\]
Here, $P^3_{ijk}$ denotes a projection operator onto the
determinant $\ket{ijk}$, while $P^3_{f_{l,m}}$,
$P^3_{f_{2k-1,m}}$, $P^3_{f_{2k,m}}$, are projection operators
onto the following functions respectively:
\begin{eqnarray}
f^3_{l,m}&=& N_{lm}\bigg( \sum^{m-1}_{i=1}\xi_i\overline\xi_m
\ket{2i-1,2i,l}-\sum^{m-1}_{i=1}|\xi_i|^2
\ket{2m-1,2m,l}\bigg),\\
&& N_{lm}=\bigg(\sum^{m-1}_{i=1}|\xi_i|^2\bigg)^{-\frac{1}{2}}
\bigg(\sum^{m}_{i=1}|\xi_i|^2\bigg)^{-\frac{1}{2}},\nonumber \\
f^3_{2k-1,m}&=&N_{km}\bigg( \sum^{m-1}_{i=1 \atop (i\ne k)}
\xi_i\overline\xi_m \ket{2i-1,2i,2k-1} -\sum^{m-1}_{i=1
\atop (i\ne k)} |\xi_i|^2\ket{2m-1,2m,2k-1}\bigg),\\
f^3_{2k,m}&=&N_{km}\bigg( \sum^{m-1}_{i=1 \atop (i\ne k)}
\xi_i\overline\xi_m \ket{2i-1,2i,2k} -\sum^{m-1}_{i=1
\atop (i\ne k)} |\xi_i|^2\ket{2m-1,2m,2k}\bigg),\\
&& N_{km}=\bigg(\sum^{m-1}_{i=1 \atop (i\ne
k)}|\xi_i|^2\bigg)^{-\frac{1}{2}} \bigg(\sum^m_{i=1 \atop (i\ne
k)}|\xi_i|^2\bigg)^{-\frac{1}{2}}.\nonumber
\end{eqnarray}
\end{theorem}
{\bf Proof.} The canonical expansion of $g^2=\sum^{s=r/2}_{i=1}
\xi_i
\sqrt{2}\,\varphi^1_{2i-1}\wedge\varphi^1_{2i}\equiv\sum^{s=r/2}_{i=1}
\xi_i\ket{2i-1,2i}$ determines the decomposition of ${\cal
H}^1={\cal R}^1\oplus \widetilde{\cal R}^1$ and the basis $\{
\varphi^1_i\}^n_{i=1}$ in ${\cal H}^1$, where the functions
$\{\varphi^1_i\}^r_{i=1}$ span the subspace ${\cal R}^1$, while
$\{\varphi^1_i\}^n_{i=r+1}$ is an orthonormal basis in the
orthogonal complement $\widetilde {\cal R}^1$ of the subspace
${\cal R}^1$ in ${\cal H}^1$. Correspondingly, the identity
operator $I^1$ on ${\cal H}^1$ has the decomposition
$I^1=\sum^r_{i=1}P^1_i+\sum^n_{i=r+1} P^1_i \equiv
P^1_{1:r}+\widetilde P^1_{1:r}$, which induces the decomposition
of the identity operator $I^{\wedge 3}$ on ${\cal H}^{\wedge 3}$
onto mutually orthogonal projection operators:
\begin{eqnarray}
I^{\wedge 3}=\big(P^1_{1:r}+\widetilde P^1_{1:r} \big)^{\wedge 3}=
\sum^3_{k=0} {3 \choose k}P^{\wedge k}_{1:r}\wedge \widetilde
P^{\wedge (3-k)}_{1:r}=\widetilde P^{\wedge 3}_{1:r}+
3P^1_{1:r}\wedge \widetilde P^{\wedge 2}_{1:r}+3P^{\wedge
2}_{1:r}\wedge \widetilde P^1_{1:r}+P^{\wedge 3}_{1:r},
\label{i3d}
\end{eqnarray}
to which in turn corresponds the decomposition of the 3-particle
antisymmetric space ${\cal H}^{\wedge 3}$ onto the mutually orthogonal
subspaces:
\begin{eqnarray}
{\cal H}^{\wedge 3}=\widetilde {\cal R}^{\wedge 3}\oplus {\cal
R}^1\wedge \widetilde{\cal R}^{\wedge 2} \oplus {\cal R}^{\wedge
2}\wedge \widetilde{\cal R}^1\oplus{\cal R}^{\wedge 3},
\label{hs3d}
\end{eqnarray}
(for proof of the above formulae see e.g. \cite{Grudzinski85},
\cite{Grudzinski87}). The subspaces on the r.h.s. of (\ref{hs3d})
are spanned by 3-particle determinants which differ between
themselves in the number of 1-particle functions belonging to
${\cal R}^1$ and $\widetilde {\cal R}^1$ (e.g. $\widetilde{\cal
R}^{\wedge 3}$ is spanned by $\{\ket{j_1,j_2,j_3}\}_j$,
$\varphi^1_j\in \widetilde{\cal R}^1$, while ${\cal R}^1\wedge
\widetilde{\cal R}^{\wedge 2}$ by $\{\ket{i,j_1,j_2}\}_{i,j}$,
$\varphi^1_i\in {\cal R}^1$, $\varphi^1_j\in \widetilde{\cal
R}^1$, etc.).
%
%
%
%

Now, comparing the spectral decomposition of the operator
$3P^2_g\wedge I^1$ (\ref{sd}) with the decomposition of the
identity operator $I^{\wedge 3}$ (\ref{i3d}) it is possible to
find all the mutually orthogonal projection operators $K^3_{0,3}$,
$K^3_{1,2}$, $K^3_{2,1}$, $K^3_{3,0}$, which constitute the
projection operator onto the kernel ${\rm Ker}\,( 3P^2_g \wedge
I^1)$ (\ref{kd0}).

\begin{itemize}
\item[$K^3_{0,3}\big)$]
$K^3_{0,3}$ denotes the projection operator onto the kernel of
$3P^2_g\wedge I^1$ which is contained in the subspace
$\widetilde{\cal R}^{\wedge 3}$ (there are no 1-particle functions from
${\cal R}^1$ in the 3-particle determinants). It is seen from
(\ref{sd}) that the whole subspace $\widetilde{\cal R}^{\wedge 3}$
is in the kernel. Hence, from (\ref{i3d}):
\[
K^3_{0,3}=\widetilde P^{\wedge 3}_{1:r}= \sum^n_{j_1,j_2,j_3=r+1}
P^1_{j_1}\wedge P^1_{j_2}\wedge P^1_{j_3}= \sum_{r+1 \leq
j_1<j_2<j_3\leq n} P^3_{j_1,j_2,j_3},
\]
where $P^1_j$ denotes the projection operator onto $\varphi
^1_j\in \widetilde{\cal R}^1$, while $P^3_{j_1,j_2,j_3}$ is the
projection operator onto the 3-particle determinant
$\ket{j_1,j_2,j_3} \in \widetilde{\cal R}^{\wedge 3}$. The
dimension of the subspace onto which projects $K^3_{0,3}:
\dim {\rm range}\, K^3_{0,3}={n-r \choose 3}={n-2s
\choose 3 }$.
\item[$K^3_{1,2}\big)$]
There are no eigenfunctions in (\ref{sd}) belonging to the
non-zero eigenvalues which are built up from determinants
containing only one function $\varphi^1_i\in {\cal R}^1$.
Therefore,
\begin{eqnarray*}
K^3_{1,2}&=&3P^1_{1:r}\wedge\widetilde P^{\wedge 2}_{1:r}=
\sum^r_{i=1}\sum_{r+1\leq j_1<j_2\leq n} 3P^1_i\wedge
P^2_{j_1,j_2}=\sum^r_{i=1}\sum_{r+1\leq j_1<j_2\leq n}
P^3_{i,j_1,j_2}=\\
&=&\sum^{s=r/2}_{i=1}\sum_{r+1\leq j_1<j_2\leq n}
\Big(P^3_{2i-1,j_1,j_2}+P^3_{2i,j_1,j_2} \Big),
\end{eqnarray*}
and  $\dim {\rm range}\, K^3_{1,2}=r{n-r \choose
2}=2s{n-2s \choose 2 }$.
\item[$K^3_{2,1}\big)$]
First, we decompose the projection operator onto the subspace
${\cal R}^{\wedge 2}\wedge \widetilde{\cal R}^1$ in the following
way:
\begin{eqnarray*}
\lefteqn{3P^{\wedge 2}_{1:r}\wedge\widetilde P^1_{1:r}=3
\left[\sum^s_{i_1=1}\Big(P^1_{2i_1-1}+P^1_{2i_1}\Big) \right]
\wedge \left[\sum^s_{i_2=1}\Big(P^1_{2i_2-1}+P^1_{2i_2}\Big)
\right]\wedge\sum^n_{j=r+1}P^1_j=\null}\\
&&\null=
3\left[\sum^s_{i_1=1}\sum^s_{i_2=1}\Big(P^1_{2i_1-1}\wedge
P^1_{2i_2-1} + P^1_{2i_1}\wedge P^1_{2i_2} + P^1_{2i_1-1}\wedge
P^1_{2i_2}+\right.\\
&&\null + \left. P^1_{2i_1}\wedge
P^1_{2i_2-1}\Big)\right]\wedge\sum^n_{j=r+1} P^1_j= \sum_{1\leq
i_1<i_2\leq s}\sum^n_{j=r+1} \Big(
P^3_{2i_1-1,2i_2-1,j}+P^3_{2i_1,2i_2,j} +\\
&&\null+
P^3_{2i_1-1,2i_2,j}+P^3_{2i_1,2i_2-1,j}\Big)+\sum^s_{i=1}\sum^n_{j=r+1}
P^3_{2i-1,2i,j}.
\end{eqnarray*}
%
%
%
%
Comparing this decomposition with (\ref{ef3}) we see that only in
the last subspace ${\rm range}\left(\,\sum^s_{i=1}\sum^n_{j=r+1}\right.$
$\left.P^3_{2i-1,2i,j}\right)$ are there eigenfunctions $g^3_l$
($l=r+1,\dots,n$) belonging to the eigenvalue different from zero.
The dimension of this subspace is $s(n-r)=s(n-2s)$. There are
$n-r$ orthonormal functions $g^3_l$. Hence, there still exists a
$s(n-r)-(n-r)=(s-1)(n-2s)$ dimensional subspace belonging to the
null-space of $3P^2_g\wedge I^1$. In this subspace we introduce
the following orthonormal basis:
\begin{eqnarray*}
f^3_{l,m}&=& N_{lm}\bigg( \sum^{m-1}_{i=1}\xi_i\overline\xi_m
\ket{2i-1,2i,l}-\sum^{m-1}_{i=1}|\xi_i|^2
\ket{2m-1,2m,l}\bigg),\\
&& N_{lm}=\bigg(\sum^{m-1}_{i=1}|\xi_i|^2\bigg)^{-\frac{1}{2}}
\bigg(\sum^{m}_{i=1}|\xi_i|^2\bigg)^{-\frac{1}{2}},\quad
l=r+1,\dots,n,\quad m=2,\dots,s.
\end{eqnarray*}
It can be checked that the functions $\{f^3_{l,m}\}$
($l=r+1,\dots,n;\quad m=2,\dots,s.$) are:
\item[$1^\circ$] orthogonal to $\{g^3_l\}$ ($l=r+1,\dots,n$),
\begin{eqnarray*}
\braket{g^3_l}{f^3_{l,m}}&=& \sum^s_{i=1}\overline\xi_i
\bra{2i-1,2i,l}N_{lm}\bigg(
\sum^{m-1}_{j=1}\xi_j\overline\xi_m
\ket{2j-1,2j,l}+\\
&-&\sum^{m-1}_{j=1}|\xi_j|^2
\ket{2m-1,2m,l}\bigg)=\\
&=&N_{lm}\bigg( \sum^{m-1}_{i=1}|\xi_i|^2\overline\xi_m
-\sum^{m-1}_{i=1}|\xi_i|^2\overline\xi_m\bigg)=0,
\end{eqnarray*}
obviously $\braket{g^3_{l_1}}{f^3_{l_2,m}}=0$ if $l_1 \ne l_2$;
\item[$2^\circ$] normalized,
\begin{eqnarray*}
\lefteqn{ \braket{f^3_{l,m}}{f^3_{l,m}}=N^2_{lm} \bigg(
\sum^{m-1}_{i=1}\overline\xi_i\xi_m \bra{2i-1,2i,l}
-\sum^{m-1}_{i=1}|\xi_i|^2
\bra{2m-1,2m,l}\bigg) \null}\\
&&\null\bigg( \sum^{m-1}_{j=1}\xi_j\overline\xi_m
\ket{2j-1,2j,l}
-\sum^{m-1}_{j=1}|\xi_j|^2 \ket{2m-1,2m,l}\bigg)=\\
&&\null=N^2_{lm}\bigg( \sum^{m-1}_{i=1}|\xi_i|^2|\xi_m|^2
+\sum^{m-1}_{i=1}|\xi_i|^2 \sum^{m-1}_{j=1}|\xi_j|^2\bigg)=\\
&&\null=N^2_{lm}\bigg( \sum^{m-1}_{i=1}|\xi_i|^2\bigg) \bigg(
|\xi_m|^2 +\sum^{m-1}_{j=1}|\xi_j|^2\bigg)= N^2_{lm}\bigg(
\sum^{m-1}_{i=1}|\xi_i|^2\bigg)\bigg(
\sum^{m}_{j=1}|\xi_j|^2\bigg)=1;
\end{eqnarray*}
\item[$3^\circ$] mutually orthogonal,
\begin{eqnarray*}
\lefteqn{ \braket{f^3_{l,m_1}}{f^3_{l,m_2}}_{m_1<m_2}=N_{lm_1}
N_{lm_2} \bigg( \sum^{m_1-1}_{i=1}\overline\xi_i\xi_{m_1}
\bra{2i-1,2i,l} -\sum^{m_1-1}_{i=1}|\xi_i|^2
\bra{2m_1-1,2m_1,l}\bigg) \null}\\
&&\null\bigg( \sum^{m_2-1}_{j=1}\xi_j\overline\xi_{m_2}
\ket{2j-1,2j,l}
-\sum^{m_2-1}_{j=1}|\xi_j|^2 \ket{2m_2-1,2m_2,l}\bigg)=\\
&&\null=N_{lm_1} N_{lm_2}\bigg(
\sum^{m_1-1}_{i=1}\overline\xi_i\xi_{m_1} \bra{2i-1,2i,l}
-\sum^{m_1-1}_{i=1}|\xi_i|^2 \bra{2m_1-1,2m_1,l}\bigg)\\
&&\null\bigg( \sum^{m_1-1}_{j=1}\xi_j\overline\xi_{m_2}
\ket{2j-1,2j,l}+
\sum^{m_2-1}_{j=m_1}\xi_j\overline\xi_{m_2}\ket{2j-1,2j,l}
-\sum^{m_2-1}_{j=1}|\xi_j|^2 \ket{2m_2-1,2m_2,l}\bigg)=\\
&&\null=N_{lm_1} N_{lm_2}\bigg(
\sum^{m_1-1}_{i=1}|\xi_i|^2\xi_{m_1}\overline\xi_{m_2}-
\sum^{m_1-1}_{i=1}|\xi_i|^2\xi_{m_1}\overline\xi_{m_2}\bigg)=0,
\end{eqnarray*}
obviously $\braket{f^3_{l_1,m}}{f^3_{l_2,m}}_{l_1 \ne l_2}=0$.\\
%
%
%
%
Thus, the sets $\{f^3_{l,m}\}$ ($l=r+1,\dots,n;\quad m=2,\dots,s$)
and $\{g^3_l\}$ ($l=r+1,\dots,n$) span the whole subspace ${\rm
range}\left(  \sum^s_{i=1}\sum^n_{j=r+1}P^3_{2i-1,2i,j}\right)$.
Now we can find the dimension of $\ker(3P^2_g\wedge I^1)$
contained in ${\cal R}^{\wedge 2}\wedge \widetilde{\cal R}^1$. It
is equal to $\dim{\rm
range}\,K^3_{2,1}=4(n-r){s\choose2}+[s(n-r)-(n-r)]=4(n-2s){s\choose2}+(s-1)(n-2s)$.
\item[$K^3_{3,0}\big)$]
$K^3_{3,0}$ is a projection operator onto the $\ker(3P^2_g\wedge
I^1)$ which is contained in the subspace ${\cal R}^{\wedge 3}$.
The dimension of the subspace ${\cal R}^{\wedge 3}$ is ${r\choose
3}={2s\choose 3}$, and there are $r=2s$ orthonormal functions
$\{g^3_{2k-1}, g^3_{2k}\}$ ($k=1,\dots,s$) belonging to the ${\rm
range}\,(3P^2_g\wedge I^1)$ in ${\cal R}^{\wedge 3}$. Hence, the
dimension of the null-space in ${\cal R}^{\wedge 3}$ is equal to
${2s \choose 3}-2s$. To find the projection operator $K^3_{3,0}$
onto this null-space, we have to decompose properly the projection
operator $P^{\wedge 3}_{1:r}$ onto the whole subspace ${\cal
R}^{\wedge 3}$. Here the calculations are longer than in the
previous cases considered above, so we give only the milestones.
\begin{eqnarray*}
\lefteqn{P^{\wedge 3}_{1:r}=\sum^s_{i_1,i_2,i_3=1}\Big(
P^1_{2i_1-1}+P^1_{2i_1}\Big)\wedge\Big(
P^1_{2i_2-1}+P^1_{2i_2}\Big)\wedge\Big(
P^1_{2i_3-1}+P^1_{2i_3}\Big)=\null}\\
&&\null=\sum_{1\leq i_1 < i_2 < i_3\leq s}\Big(
P^3_{2i_1-1,2i_2-1,2i_3-1}+P^3_{2i_1-1,2i_2-1,2i_3}+
P^3_{2i_1-1,2i_2,2i_3-1}+\\
&&\null+P^3_{2i_1,2i_2-1,2i_3-1} +
P^3_{2i_1-1,2i_2,2i_3}+P^3_{2i_1,2i_2-1,2i_3}+
P^3_{2i_1,2i_2,2i_3-1}+P^3_{2i_1,2i_2,2i_3}\Big)+\\
&&\null+\sum^s_{i_1,i_2=1\atop (i_1\ne i_2)}\Big(
P^3_{2i_1-1,2i_1,2i_2-1}+P^3_{2i_1-1,2i_1,2i_2}\Big).
\end{eqnarray*}
Here we have used the facts that:
\item[$1^\circ$]
\begin{eqnarray*}
\sum^s_{i_1,i_2,i_3=1}P^1_{2i_1-1}\wedge P^1_{2i_2-1}\wedge
P^1_{2i_3-1}=\sum_{1\leq i_1 < i_2 < i_3\leq s} 3!\,
P^1_{2i_1-1}\wedge P^1_{2i_2-1}\wedge P^1_{2i_3-1}=\\
=\sum_{1\leq i_1 < i_2 < i_3\leq s}P^3_{2i_1-1,2i_2-1,2i_3-1},
\end{eqnarray*}
and similarly for the projector with three even indices;
\item[$2^\circ$]
\begin{eqnarray*}
\lefteqn{\sum^s_{i_1,i_2,i_3=1}\Big(P^1_{2i_1-1}\wedge
P^1_{2i_2-1}\wedge P^1_{2i_3}+P^1_{2i_1-1}\wedge P^1_{2i_2}\wedge
P^1_{2i_3-1}+ P^1_{2i_1}\wedge P^1_{2i_2-1}\wedge
P^1_{2i_3-1}\Big)=\null}\\
&&\null=\sum^s_{i_1,i_2=1}
P^1_{2i_1-1}\wedge P^1_{2i_2-1}\wedge P^1_{2i_1}+
\sum^s_{i_1,i_2=1}
P^1_{2i_1-1}\wedge P^1_{2i_2-1}\wedge P^1_{2i_2}+ \\
&&\null+\sum^s_{i_1,i_2,i_3=1 \atop (i_1\ne i_2\ne i_3)}
P^1_{2i_1-1}\wedge P^1_{2i_2-1}\wedge P^1_{2i_3}+
\sum^s_{i_1,i_3=1}
P^1_{2i_1-1}\wedge P^1_{2i_1}\wedge P^1_{2i_3-1}+\\
&&\null+\sum^s_{i_1,i_2=1}
P^1_{2i_1-1}\wedge P^1_{2i_2}\wedge P^1_{2i_2-1}+
\sum^s_{i_1,i_2,i_3=1 \atop (i_1\ne i_2\ne i_3)}
P^1_{2i_1-1}\wedge P^1_{2i_2}\wedge P^1_{2i_3-1}+\\
&&\null+\sum^s_{i_1,i_3=1}
P^1_{2i_1}\wedge P^1_{2i_1-1}\wedge P^1_{2i_3-1}+
\sum^s_{i_1,i_2=1}
P^1_{2i_1}\wedge P^1_{2i_2-1}\wedge P^1_{2i_1-1}+ \\
&&\null+\sum^s_{i_1,i_2,i_3=1 \atop (i_1\ne i_2\ne i_3)}
P^1_{2i_1}\wedge P^1_{2i_2-1}\wedge P^1_{2i_3-1}=
%
%
%
%
\sum^s_{i_1,i_2=1 \atop (i_1\ne i_2)}3!\,
P^1_{2i_1-1}\wedge P^1_{2i_1}\wedge P^1_{2i_2-1}+\\
&&\null+\sum_{1\leq i_1<i_2<i_3 \leq s}3!\, \Big(
P^1_{2i_1-1}\wedge P^1_{2i_2-1}\wedge P^1_{2i_3}+
P^1_{2i_1-1}\wedge P^1_{2i_2}\wedge P^1_{2i_3-1}+
P^1_{2i_1}\wedge P^1_{2i_2-1}\wedge P^1_{2i_3-1}\Big)=\\
&&\null=\sum^s_{i_1,i_2=1 \atop i_1\ne i_2}
P^3_{2i_1-1,2i_1,2i_2-1}+
\sum_{1\leq i_1<i_2<i_3 \leq s}\Big(
P^3_{2i_1-1,2i_2-1,2i_3}+
P^3_{2i_1-1,2i_2,2i_3-1}+
P^3_{2i_1,2i_2-1,2i_3-1}\Big).
\end{eqnarray*}
Similarly we proceed with the projectors containing two even and one odd indices.
\end{itemize}

It follows from (\ref{sd}) that only in the subspace ${\rm
range}\,\sum^s_{i_1,i_2=1\atop (i_1\ne i_2)}
\Big(P^3_{2i_1-1,2i_1,2i_2-1}+P^3_{2i_1-1,2i_1,2i_2}\Big)$
are there  $r=2s$ eigenfunctions $\{g^3_{2k-1},g^3_{2k}\}$
($k=1,\dots,s=r/2$) belonging to non-zero eigenvalues of
$3P^2_g\wedge I^1$. Since this subspace is $2s(s-1)$ dimensional,
there still exists  in it the $2s(s-1)-2s=2s(s-2)$ dimensional
null-space of $3P^2_g\wedge I^1$. Within this subspace we can 
establish the following orthonormal basis:
\begin{eqnarray*}
f^3_{2k-1,m}&=&N_{km}\bigg( \sum^{m-1}_{i=1 \atop (i\ne k)}
\xi_i\overline\xi_m \ket{2i-1,2i,2k-1} -\sum^{m-1}_{i=1
\atop (i\ne k)} |\xi_i|^2\ket{2m-1,2m,2k-1}\bigg),\\
f^3_{2k,m}&=&N_{km}\bigg( \sum^{m-1}_{i=1 \atop (i\ne k)}
\xi_i\overline\xi_m \ket{2i-1,2i,2k} -\sum^{m-1}_{i=1
\atop (i\ne k)} |\xi_i|^2\ket{2m-1,2m,2k}\bigg),\\
N_{km}&=&\bigg(\sum^{m-1}_{i=1 \atop (i\ne
k)}|\xi_i|^2\bigg)^{-\frac{1}{2}} \bigg(\sum^m_{i=1 \atop
(i\ne k)}|\xi_i|^2\bigg)^{-\frac{1}{2}},\\
&& k=1,\dots,s;\quad m\in J, \quad
J=\left\{
\begin{array}{l@{\quad \mbox{for} \quad}l}
\{3,\dots,s\},& k=1\\
\{2,\dots,k-1,k+1,\dots,s\},& k=2,\dots,s.
\end{array}\right.
\end{eqnarray*}
The orthogonality between any function with "odd index $k$" to any
one with "even index $k$" as well as between functions with
different indices $k$ is seen by inspection. The normalization,
and other orthogonality relations can be proved in a way similar
to that used in the previous case $K^3_{2,1}$, and we omit these
calculations for brevity. Now, we can calculate the dimension of
the $\ker(3P^2_g\wedge I^1)$ contained in ${\cal R}^{\wedge 3}$:
it is equal to $\dim{\rm range}\,K^3_{3,0}=8{s\choose
3}+2s(s-2)$. Thus, in principle the proof is done. To make sure
that everything that belongs to the $\ker(3P^2_g\wedge I^1)$ is
taken into account, we can perform the dimensionality test. The
subspace ${\cal H}^{\wedge 3}$ is ${n \choose 3}$ dimensional,
there are $n$ orthonormal eigenfunctions belonging to the non-zero
eigenvalues of the operator $3P^2_g\wedge I^1$, therefore the
dimension of $\ker(3P^2_g\wedge I^1)$ is ${n\choose 3}-n$. On the
other hand this should be equal to
\begin{eqnarray*}
\lefteqn{\sum^3_{i=0}\dim {\rm range\,}
K^3_{i,3-i}=\null}\\
&&\null={n-2s \choose 3}+2s{n-2s\choose 2}+\left[4(n-2s){s\choose
2}+(s-1)(n-2s)\right]+ \left[8{s\choose 3}+2s(s-2)\right].
\end{eqnarray*}
This sum is really equal to ${n\choose 3}-n$, and this completes
the proof.

\end{document}